\documentclass[aps,prb,twocolumn]{revtex4}
\usepackage{amsfonts}
\usepackage{amsmath}
\usepackage{graphicx}
\usepackage{dsfont}
\usepackage{diagbox}
\usepackage{array}
\usepackage{booktabs}
\usepackage{makecell}
\usepackage{bbm}
\usepackage{amssymb}

\newcommand{\PreserveBackslash}[1]{\let\temp=\\#1\let\\=\temp}
\newcolumntype{C}[1]{>{\PreserveBackslash\centering}p{#1}}
\newcolumntype{R}[1]{>{\PreserveBackslash\raggedleft}p{#1}}
\newcolumntype{L}[1]{>{\PreserveBackslash\raggedright}p{#1}}

\input{tcilatex}

\begin{document}

\title{Classifying parafermionic gapped phases using matrix product states}
\author{Wen-Tao Xu$^{1}$ and Guang-Ming Zhang$^{1,2}$}
\affiliation{$^{1}$State Key Laboratory of Low-Dimensional Quantum Physics and Department
of Physics, Tsinghua University, Beijing 100084, China. \\
$^{2}$Collaborative Innovation Center of Quantum Matter, Beijing 100084,
China.}
\date{\today }

\begin{abstract}
In the Fock representation, we construct matrix product states (MPS) for
one-dimensional gapped phases for $\mathbb{Z}_{p}$ parafermions. From the
analysis of irreducibility of MPS, we classify all possible gapped phases of 
$\mathbb{Z}_{p}$ parafermions without extra symmetry other than $\mathbb{Z}%
_{p}$ charge symmetry, including topological phases, spontaneous symmetry
breaking phases and a trivial phase. For all phases, we find the irreducible
forms of local matrices of MPS, which span different kinds of graded
algebras. The topological phases are characterized by the non-trivial simple 
$\mathbb{Z}_{p}$ graded algebras with the characteristic graded centers,
yielding the degeneracies of the full transfer matrix spectra uniquely. But
the spontaneous symmetry breaking phases correspond to the trivial
semisimple $\mathbb{Z}_{p/n}$ graded algebras, which can be further reduced
to the trivial simple $\mathbb{Z}_{p/n}$ graded algebras, where $n$ is the
divisor of $p$. So the present results provide the complete classification
of the parafermionic gapped phases and deepen our understanding of
topological phases in one dimension.
\end{abstract}

\maketitle

\section{Introduction}

Much of the extensive research in the last few years has focused on
topological phases of matter and their classifications\cite%
{Schnyder2008,Kitaev2009,Ryu2010}. A prime example is the one-dimensional
topological superconducting phase of paired fermions, which is characterized
by Majorana zero modes at the edges\cite{Kitaev2001}. Such zero modes
actually resemble the ones found in the cores of vortices in two-dimensional
topological superconductors\cite{ReadGreen}, and have been shown useful for
quantum information processing\cite{AliceaRPP}. However, it has been argued
that one-dimensional fermion systems with interactions and no extra symmetry
than the intrinsic fermion parity can only realize two topologically
distinct phases\cite{Fidkowski2011,Turner2011}. In order to search for a
universal quantum computation platform and fully understand topological
excitations in strongly interacting electron systems, more exotic
parafermion excitations have been investigated in an effectively
one-dimensional system\cite{Fendley2012,Pollmann2013,AliceaFendley}, which
exists at the edges of a two-dimensional fractionalized topological state
and cannot be realized in a strictly one-dimensional system\cite%
{Lindner,Clarke,Cheng}.

For various correlated low-dimensional gapped systems, it has been known
that matrix product states\cite{FannesWerner1992} (MPS) and their high
dimensional generalizations, tensor network states or projective entangled
pair states\cite{VerstraeteCirac} (PEPS), have been proven increasingly
successful. The framework of MPS and PEPS naturally provides an efficient
method to classify topologically ordered phases\cite{Schuch2013,Haegeman2015}%
, symmetry protected topological phases\cite%
{Chen-Gu-Wen-2011,Schuch,chen-gu-liu-wen}, and the long-range ordered phases
with spontaneous symmetry breaking\cite{Chen-Gu-Wen-2011,Rispler2015}.
However, it is not straight forward to extend the MPS representation to the
one-dimensional parafermion systems. Recently, the fermionic MPS have been
successfully constructed by using the language of super vector space, and
all possible topological phases with additional symmetries in terms of
Majorana fermions have been classified within the matrix product
representation\cite{Bultinck2017,Kapustin2016}. By generalizing the concepts
of fermion parity and associated Fock space, the present authors have
proposed a general framework to construct the MPS of $\mathbb{Z}_{3}$
parafermions in the Fock representation, and the corresponding parent
Hamiltonians have been also derived\cite{XuZhang}. Therefore, the road has
been cleared to classify all possible $\mathbb{Z}_{p}$ parafermion gapped
phases within the framework of the MPS.

In this paper, we first review the Fock space of parafermions and then
construct the parafermionic MPS. From the analysis of irreducibility of
these parafermionic MPS, we provide the complete classification of all
possible gapped phases without extra symmetry, compared to the previous
classification based on the edge fractionalization\cite{Pollmann2013,Quella}%
. More importantly, we find that the various irreducible forms for local
matrices of MPS spanned different kinds of graded algebras characterize
distinct parafermionic gapped phases. The local matrices of MPS describing
topological phases span the non-trivial simple $\mathbb{Z}_{p}$ graded
algebras with characteristic centers, resulting in the degeneracies of the
full transfer matrix spectra and entanglement spectra in the thermodynamic
limit. Meanwhile, the spontaneous symmetry breaking phases correspond to the
trivial semisimple $\mathbb{Z}_{p/n}$ graded algebras ($n$ is a divisor of $%
p $), which can be further reduced to the trivial simple $\mathbb{Z}_{p/n}$
graded algebras. The trivial phase corresponds to trivial simple $\mathbb{Z}%
_{p}$ graded algebra without non-trivial center. Furthermore, we also found
that the topological order is closely related to the non-trivial center of
the graded algebra, giving rise to the degeneracy of the full transfer
matrix spectrum and the existence of parafermion edge zero modes.

In Sec. II, we briefly review the Fock space of parafermions and present the
construction of the $\mathbb{Z}_{p}$ parafermionic MPS. Then, we outline the
general classification framework and the detailed classifications for $%
\mathbb{Z}_{3}$, $\mathbb{Z}_{4}$, $\mathbb{Z}_{6}$, and $\mathbb{Z}_{8}$
parafermionic gapped phases are considered separately in Sec. III. In Sec.
IV, the general irreducible forms for various phases of the $\mathbb{Z}_{p}$
parafermions are summarized, and the topological order in the form of MPS is
discussed. Conclusion and outlook are given in Sec. V. The related concepts
used in the classification scheme are listed in the appendix.

\section{Parafermionic MPS}

\subsection{Fock space of parafermions}

In order to keep our discussion self-contained, we first briefly review the
parafermions and their Fock space. It has been known that, from the $\mathbb{%
Z}_{p}$ spin degrees of freedom of the clock models, the parafermions are
defined by a generalized Jordan-Wigner transformation as\cite%
{FradkinKadonoff,AlcarazKoberle} 
\begin{equation}
\chi _{2l-1}=\left( \prod_{k<l}\tau _{k}\right) \sigma _{l},\quad \chi
_{2l}=-e^{i\pi /p}\left( \prod_{k\leq l}\tau _{k}\right) \sigma _{l},
\end{equation}%
where the $\mathbb{Z}_{p}$ spin operators satisfy the following relations 
\begin{equation}
\sigma _{l}^{p}=\tau _{l}^{p}=1,\quad \sigma _{l}\tau _{m}=\omega
_{p}^{\delta _{l,m}}\tau _{m}\sigma _{l},
\end{equation}%
with $\omega _{p}=e^{i2\pi /p}$. So the algebras of the parafermions are 
\begin{equation}
\chi _{l}^{p}=1,\quad \chi _{l}^{p-1}=\chi _{l}^{\dagger },\quad \chi
_{l}\chi _{m}=\omega _{p}\chi _{m}\chi _{l},
\end{equation}%
for $l<m$. These are the generalized Clifford $\mathbb{Z}_{p}$ graded
algebras, and the parafermions are referred to as the Weyl parafermions,
because it was first introduced by Weyl\cite{Weyl1950}. It has been noticed
that the second quantized description of the Weyl parafermions is given by
the Fock parafermions\cite{CobaneraOrtiz}. So the basis of $\mathbb{Z}_{p}$
Fock parafermions can be assumed as $|i_{1}i_{2}\cdots i_{L}\rangle $, where 
$i_{1}$, $i_{2}$, $\cdots $, $i_{L}$ $\in \mathbb{Z}_{p}\equiv \left\{
0,1,2,\cdots ,p-1\right\} $ are the respective occupation numbers of the
single particle orbitals. The general structure of the Fock space is defined
by 
\begin{equation}
\mathbb{V}_{F}=\bigoplus_{M=0}^{L(p-1)}\text{Span}\left\{ |i_{1}i_{2}\cdots
i_{L}\rangle ,|\sum_{l=1}^{L}i_{l}=M\right\} .
\end{equation}%
In the following the abbreviated notation $|i_{l}\rangle =|0\cdots
i_{l}\cdots 0\rangle $ denotes the single-particle states. In order to
encode the parafermion statistics into the Fock space, the graded tensor
product $\otimes _{g}$ building many-body states is introduced as 
\begin{eqnarray}
\langle i_{1}i_{2}\cdots i_{L}| &=&\langle i_{L}|\otimes _{g}\cdots \otimes
_{g}\langle i_{2}|\otimes _{g}\langle i_{1}|,  \notag \\
|i_{1}i_{2}\cdots i_{L}\rangle &=&|i_{1}\rangle \otimes _{g}|i_{2}\rangle
\otimes _{g}\cdots \otimes _{g}|i_{L}\rangle ,
\end{eqnarray}%
which describes the graded structure of Hilbert space mathematically. The
crucial ingredient of the graded tensor product is the following isomorphism
mapping $\mathcal{F}$: 
\begin{eqnarray}
\mathcal{F}(|i_{l}\rangle \otimes _{g}|j_{m}\rangle ) &\equiv &\omega
_{p}^{ij}|j_{m}\rangle \otimes _{g}|i_{l}\rangle ,  \notag \\
\mathcal{F}(\langle i_{l}|\otimes _{g}|j_{m}\rangle ) &\equiv &\bar{\omega}%
_{p}^{ij}|j_{m}\rangle \otimes _{g}\langle i_{l}|,
\end{eqnarray}%
for $l<m$. The isomorphism $\mathcal{F}$ exchanges two nearby local Fock
states, and the whole Fock space is a graded vector space, which is a
generalization of super vector space of the fermions\cite{Bultinck2017}.
Thus parafermion statistics is encoded into the Fock space by the
isomorphism, which becomes crucial for the construction of the MPS wave
functions.

Since a contraction is necessary for tensor networks, the homomorphism $%
\mathcal{C}$ has to be defined via a mapping $\mathbb{V}_{F}^{\ast }\otimes
_{g}\mathbb{V}_{F}\rightarrow \mathbb{C}$: 
\begin{equation}
\mathcal{C}\left( \langle i_{l}|\otimes _{g}|j_{l}\rangle \right) =\langle
i_{l}|j_{l}\rangle =\delta _{i_{l},j_{l}},
\end{equation}%
which is nothing but the inner product and orthonormal. From the above
propositions, the $p$-exclusion principle can be derived as 
\begin{equation}
(|i_{l}=1\rangle )^{\otimes _{g}p}\equiv |i_{l}=p\rangle =0.
\end{equation}%
So the dimension of the Fock space of parafermions is determined as $p^{L}$.
The creation and annihilation operators of Fock space can also be
introduced, and their commutation relations have been derived\cite%
{CobaneraOrtiz}.

Furthermore, the local charge operator can be defined by $\mathbf{Q}%
_{l}=-e^{i\pi /p}\chi _{2l-1}^{\dagger }\chi _{2l}$, and the global one is
accordingly given by $\mathbf{Q}=\prod_{l}\mathbf{Q}_{l}$, determining the
charge of the Fock basis as\cite{XuZhang} 
\begin{equation}
\mathbf{Q}|I\rangle =\mathbf{Q}|i_{1}i_{2}\cdots i_{L}\rangle =\omega
_{p}^{\sum_{i_{l}=1}^{L}i_{l}}|i_{1}i_{2}\cdots i_{L}\rangle .
\end{equation}%
Then the charge of the Fock state $|I\rangle $ can be calculated as $%
|I|=(\sum_{i_{l}=1}^{L}i_{l}$) $\bmod$ $p$, while the charge of the bar $%
\langle I|$ is given by $-|I|$. It should be emphasized that only the
many-body states which are superpositions of the Fock states with the same
charge have well-defined charges.

\subsection{MPS for $\mathbb{Z}_{p}$ parafermions}

To construct the MPS with physical degrees of freedom of dimension $d$, we
have to introduce two auxiliary virtual degrees of freedom of dimension $D$.
Two virtual degrees of freedom form a maximally entangled state on the
neighboring sites, while the virtual degrees of freedom on the same site are
mapped to the physical degree of freedom. In the Fock space of parafermions,
we can write down the local tensor as 
\begin{equation}
\mathbf{A}[l]=\sum_{\alpha \beta i}A[l]_{\alpha \beta }^{[i]}|\alpha
_{l})\otimes _{g}|i_{l}\rangle \otimes _{g}(\beta _{l+1}|,
\label{LocalTensor}
\end{equation}%
where $\mathbf{A}[l]\in \mathbb{V}_{l}\otimes _{g}\mathbb{H}_{l}\otimes _{g}%
\mathbb{V}_{l+1}^{\ast }$, $l$ denotes the site index, $|i_{l}\rangle $
stands for the physical state with the charge $|i|\in \mathbb{Z}_{p}$, and $%
|\alpha _{l})$, $(\beta _{l+1}|$ stand for the virtual states with the
charges $|\alpha |$, $-|\beta |\in \mathbb{Z}_{p}$ respectively.

Since the charge symmetry acts locally on the tensor networks, we impose the
constraint that all the local tensors $\mathbf{A}[l]$\ must have
well-defined charges. This enforces that the local matrices $\mathbf{A}%
^{[i]}[l]$ as the components of local tensors have well-defined charges as
well. Then we choose the simplest convention that all local tensors $\mathbf{%
A}[l]$ are charge-0, so that the different orders of the tensors $\mathbf{A}%
[l]$ does not induce any phases and the total charges of the tensor networks
are independent of the system size\cite{XuZhang}. The charges of the local
matrices $\mathbf{A}^{[i]}=\sum_{\alpha \beta }A_{\alpha \beta
}^{[i]}|\alpha _{l})\otimes _{g}(\beta _{l+1}|$ are given by $(\alpha -\beta
)$ $\mathrm{mod}$ $p$, shown in Table. \ref{table1}. To ensure that the
local tensors are charge-0, the matrices $A^{[i]}$ must have the following
block structures under the basis with the well-defined $\mathbb{Z}_{p}$
charges: 
\begin{eqnarray}  \label{standradform}
A^{[i]} &=&\left[ 
\begin{array}{ccccc}
a_{0}^{[i]} & 0 & 0 & \cdots  & 0 \\ 
0 & a_{1}^{[i]} & 0 & \cdots  & 0 \\ 
0 & 0 & a_{2}^{[i]} & \cdots  & 0 \\ 
\vdots  & \vdots  & \vdots  & \ddots  & 0 \\ 
0 & 0 & 0 & 0 & a_{p-1}^{[i]}%
\end{array}%
\right] ,|i|=0;  \notag  \label{block} \\
A^{[i]} &=&\left[ 
\begin{array}{ccccc}
0 & a_{0}^{[i]} & 0 & \cdots  & 0 \\ 
0 & 0 & a_{1}^{[i]} & \cdots  & 0 \\ 
0 & 0 & 0 & \cdots  & 0 \\ 
\vdots  & \vdots  & \vdots  & \ddots  & \vdots  \\ 
a_{p-1}^{[i]} & 0 & 0 & 0 & 0%
\end{array}%
\right] ,|i|=1;  \notag \\
&\vdots &\quad   \notag \\
A^{[i]} &=&\left[ 
\begin{array}{ccccc}
0 & 0 & 0 & \cdots  & a_{0}^{[i]} \\ 
a_{1}^{[i]} & 0 & 0 & \cdots  & 0 \\ 
0 & a_{2}^{[i]} & 0 & \cdots  & 0 \\ 
\vdots  & \vdots  & \vdots  & \ddots  & \vdots  \\ 
0 & 0 & 0 & a_{p-1}^{[i]} & 0%
\end{array}%
\right] ,|i|=p-1,
\end{eqnarray}%
where $a_{r}^{[i]}$ with $r\in \mathbb{Z}_{p}$ are matrices in the
sub-blocks with smaller virtual dimensions. Actually the structures of the
local matrices are determined by the fact that the Fock space is a graded
vector space. Moreover, the charges of the local matrices can be revealed by
the representation of charge operator $Q_{p}=$ diag$\left( 1,\omega
_{p},\omega _{p}^{2},\cdots ,\omega _{p}^{p-1}\right) $ as 
\begin{equation}
\left( Q_{p}\otimes \mathbbm{1}\right) ^{-1}A^{[i]}\left( Q_{p}\otimes %
\mathbbm{1}\right) =\omega _{p}^{|i|}A^{[i]}.
\end{equation}%
For the later discussion, it is useful to introduce another $p\times p$
matrix 
\begin{equation}
Y_{p}=\left[ 
\begin{array}{ccccc}
0 & 1 & 0 & \cdots  & 0 \\ 
0 & 0 & 1 & \cdots  & 0 \\ 
0 & 0 & 0 & \cdots  & 0 \\ 
\vdots  & \vdots  & \vdots  & \ddots  & 1 \\ 
1 & 0 & 0 & 0 & 0%
\end{array}%
\right] ,  \label{Y}
\end{equation}%
as the regular representation of generator of $\mathbb{Z}_{p}$ symmetry. It
will be frequently used and played a significant role in the following
classification. 
\begin{table}[tbp]
\caption{The charges of elements of the local tensor $\mathbf{A}^{[i]}$}
\label{table1}%
\begin{tabular}{|C{2cm}|C{0.8cm}|C{0.8cm}|C{0.8cm}|C{0.8cm}|C{0.8cm}|C{0.8cm}|}
\hline
\diagbox{$|\alpha|$}{$|\mathbf{A}^{[i]}_{\alpha\beta}{[l]}|$}{-$|\beta|$}& 0 &  $p-1$& $p-2$ &$p-3$ &$\cdots$& 1  \\ \hline
0 &  0 & $p-1$& $p-2$ &$p-3$ &$\cdots$&1 \\ \hline
1 &  1 & 0 &  $p-1$&$p-2$ &$\cdots$ &2\\ \hline
2 &  2 & 1 & 0&$p-1$ & $\cdots$&3\\ \hline
3 &  3 & 2 & 1&$0$ & $\cdots$&4\\ \hline
$\vdots$ & $\vdots$ & $\vdots$ & $\vdots$&$\vdots$&$\ddots$&\vdots \\ \hline
$p-1$ & $p-1$ & $p-2$ & $p-3$&$p-4$&$\cdots$&0 \\ \hline
\end{tabular}
\end{table}

So the parafermionic MPS can be constructed by taking graded tensor product
of local tensors and contracting the virtual bonds between nearby local
tensors with the homomorphism $\mathcal{C}$. The contraction does not affect
the charges of the tensor networks, because the charge of $|\alpha
_{l})\otimes _{g}(\alpha _{l}|$ is zero. Therefore, the general
parafermionic MPS is expressed as 
\begin{eqnarray}
|\psi \rangle  &=&\mathcal{C}(\mathbf{C}_{a}\otimes _{g}\mathbf{A}[1]\otimes
_{g}\mathbf{A}[2]\otimes _{g}\cdots \otimes _{g}\mathbf{A}[L])  \notag \\
&=&\sum_{i_{1}..i_{N}}\left( C_{a}^{T}A^{[i_{1}]}\cdots A^{[i_{L}]}\right)
|i_{1}\cdots i_{L}\rangle ,
\end{eqnarray}%
where $\mathbf{C}_{a}=\sum_{\gamma \delta }C_{a,\gamma \delta }(\gamma
_{1}|\otimes _{g}|\delta _{L})$ is the closure tensor and different choices
of $\mathbf{C}_{a}$ just result in the different charges of the closed MPS
wave functions\cite{XuZhang}. It should be emphasized that, unlike the
Majorana fermion chains, the periodic boundary condition can not reconcile
with the algebras of parafermions, so the periodic boundary condition for
parafermion chains does not exist. How to define the Hamiltonian for closed
boundary conditions has been specifically discussed\cite%
{XuZhang,Alexandradinata}.

\section{Classification of $\mathbb{Z}_{p}$ parafermionic MPS with
irreducibility}

Irreducibility is the most important property for a general MPS, because the
irreducible MPS determines the major physical properties of the system.
Irreducible forms of bosonic and fermionic MPS have been constructed, the
concept of irreducibility of fermionic MPS is quite different from that of
bosonic MPS. For the fermionic MPS, there are two types of irreducible
fermionic MPS\cite{Bultinck2017,Fidkowski2011}, one is called even type with
the local matrices with the block structures: 
\begin{equation}
A^{[i]}=\left[ 
\begin{array}{cc}
a_{0}^{[i]} & 0 \\ 
0 & a_{1}^{[i]}%
\end{array}%
\right] ,|i|=0;\text{ }A^{[i]}=\left[ 
\begin{array}{cc}
0 & a_{0}^{[i]} \\ 
a_{1}^{[i]} & 0%
\end{array}%
\right] ,|i|=1,  \label{evenZ2}
\end{equation}%
where the sub-block matrices can \textit{not} be equal under gauge
transformations. These irreducible matrices span the even type simple $%
\mathbb{Z}_{2}$ graded matrix algebra with the center consisting of
multiples of the identity. So it is as simple as the ungraded algebra. While
the other is called the odd type which can be gauge transformed into 
\begin{equation}
A^{[i]}=\left[ 
\begin{array}{cc}
a^{[i]} & 0 \\ 
0 & a^{[i]}%
\end{array}%
\right] ,|i|=0;\text{ }A^{[i]}=\left[ 
\begin{array}{cc}
0 & a^{[i]} \\ 
a^{[i]} & 0%
\end{array}%
\right] ,|i|=1.  \label{oddZ2}
\end{equation}%
Then we can be further expressed into a more compact form: 
\begin{equation}
A^{[i]}=Y_{2}^{|i|}\otimes a^{[i]},\text{ }|i|=0,1,
\end{equation}%
where $Y_{2}$ has been defined in Eq. (\ref{Y}). Thus $A^{[i]}$ of
irreducible MPS span an odd type simple $\mathbb{Z}_{2}$ graded algebra with
the center consisting of multiples of $\mathbbm{1}$ and $Y_{2}$,
characterizing the non-trivial topological phase with unpaired Majorana zero
modes at the edges of one-dimensional systems. These two types of
irreducible MPS represent the $\mathbb{Z}_{2}$ classification of
one-dimensional interacting fermionic systems\cite{Fidkowski2011,Turner2011}.

Expanding the analysis of irreducibility, we can derive all irreducible $%
\mathbb{Z}_{p}$ parafermionic MPS, and the different algebras spanned by the
local matrices correspond to distinct gapped phases. Therefore, we can
establish a complete classification of all one-dimensional parafermionic
gapped phases, including the topological phases and conventional spontaneous
symmetry breaking phases. The symmetry protected topological phases are not
included, because no extra symmetry than the parafermionic charge is
involved and $\mathbb{Z}_{p}$ symmetry is not sufficient to support symmetry
protected topological phases. The general method is as follows. We first
assume that all $A^{[i]}$ have an irreducible invariant subspace with the
corresponding orthogonal projector $P_{0}$. Then we analyze all the
commutation relations between the operators $P_{0}$ and $Q_{p}^{r}$, where $%
r $ is a positive integer. $P_{0}$ and $Q_{p}^{r}$ generate a space which
contains all $\mathbb{Z}_{p}$ charge sectors. Finally, by using the
invariant subspace projectors containing all charge sectors, we can
determine the irreducible structures of $A^{[i]}$ and the corresponding
algebras spanned by them. The different algebras associate with different
gapped phases. In the following, several important cases are carried out in
detail.

\subsection{Irreducibility of $\mathbb{Z}_{3}$ parafermion MPS}

Now we first derive two types of irreducible MPS for $\mathbb{Z}_{3}$
parafermion chains\cite{XuZhang}. Without the lost of generality, we assume
that there is an irreducible invariant subspace projector $P_{0}$ of local
matrices $A^{[i]}$, i.e. 
\begin{equation}
A^{[i]}P_{0}=P_{0}A^{[i]}P_{0}.
\end{equation}%
Because $A^{[i]}$ have a definite charge, $Q_{3}^{-1}A^{[i]}Q_{3}=\omega
^{|i|}A^{[i]}$, where $Q_{3}$ is the $\mathbb{Z}_{3}$ charge matrix, we can
further derive 
\begin{equation}
A^{[i]}P_{1}=P_{1}A^{[i]}P_{1},A^{[i]}P_{2}=P_{2}A^{[i]}P_{2},  \notag
\end{equation}%
where $P_{1}=Q_{3}P_{0}Q_{3}^{-1}$ and $P_{2}=Q_{3}^{-1}P_{0}Q_{3}$ are also
invariant subspace projectors. Since $P_{0}$ is already associated with an
irreducible invariant space, $P_{0}$, $P_{1}$ and $P_{2}$ must be either the
same or mutually orthogonal, otherwise it will contradict with the fact that 
$P_{0}$ is already associated with an irreducible invariant space. So there
are two different situations, we discuss them separately.

1. $\left[ P_{0},Q_{3}\right] =0$

It can be simply determined that $P_{0}=P_{1}=P_{2}$, and $P_{0}$ contains
all three charge sectors, indicating that $A^{[i]}$ in $P_{0}$ irreducible
invariant subspace preserve the $\mathbb{Z}_{3}$ charge symmetry. Thus all $%
A^{[i]}$ in this invariant subspace will have the initial structures shown
in Eq. (\ref{block}) and span a trivial simple $\mathbb{Z}_{3}$ graded
algebra.

2. $\left[ P_{0},Q_{3}\right] \neq 0$

In this situation, $P_{0}$, $P_{1}$ and $P_{2}$ are mutually orthogonal
projectors, the corresponding invariant subspaces do not contain all $%
\mathbb{Z}_{3}$ charge sectors. For the parafermionic MPS, the $\mathbb{Z}%
_{3}$ charge symmetry can never be broken and the MPS can not be reduced,
since the invariant spaces do not contain all charge sectors. The reduced
matrices also break the $\mathbb{Z}_{3}$ graded structures of local matrices
and no longer span a $\mathbb{Z}_{3}$ graded algebra. Thus, the concept of
irreducibility should be reformulated. Notice that $%
[P_{0}+P_{1}+P_{2},Q_{3}]=0$, and the total invariant space is the complete,
leading to $P_{0}+P_{1}+P_{2}=\mathbbm{1}$. The idempotency requires $%
P_{0}^{2}=P_{0}$, $P_{1}^{2}=P_{1}$ and $P_{2}^{2}=P_{2}$. From these
constraints, the invariant subspace projectors can be derived as 
\begin{equation*}
P_{0}=\frac{1}{3}\left[ 
\begin{array}{ccc}
\mathbbm{1} & U_{1} & U_{1}U_{2} \\ 
U_{1}^{\dagger } & \mathbbm{1} & U_{2} \\ 
U_{2}^{\dagger }U_{1}^{\dagger } & U_{2}^{\dagger } & \mathbbm{1}%
\end{array}%
\right] ,
\end{equation*}%
where $U_{1}$ and $U_{2}$ are unitary block matrices with the same
dimensions. Since $A^{[i]}P_{j}=P_{j}A^{[i]}P_{j}$, we can obtain 
\begin{eqnarray}
A^{[i]} &=&\left[ 
\begin{array}{ccc}
a_{0}^{[i]} & 0 & 0 \\ 
0 & U_{1}^{\dagger }a_{0}^{[i]}U_{1} & 0 \\ 
0 & 0 & U_{2}^{\dagger }U_{1}^{\dagger }a_{0}^{[i]}U_{1}U_{2}%
\end{array}%
\right] ,|i|=0;  \notag  \label{Z3nontrivial} \\
\text{ }A^{[i]} &=&\left[ 
\begin{array}{ccc}
0 & a_{0}^{[i]} & 0 \\ 
0 & 0 & U_{1}^{\dagger }a_{0}^{[i]}U_{2} \\ 
U_{2}^{\dagger }U_{1}^{\dagger }a_{0}^{[i]}U_{1}^{\dagger } & 0 & 0%
\end{array}%
\right] ,|i|=1;  \notag \\
\text{ }A^{[i]} &=&\left[ 
\begin{array}{ccc}
0 & 0 & a_{0}^{[i]} \\ 
U_{1}^{\dagger }a_{0}^{[i]}U_{2}^{\dagger }U_{1}^{\dagger } & 0 & 0 \\ 
0 & U_{2}^{\dagger }U_{1}^{\dagger }a_{0}^{[i]]}U_{2}^{\dagger } & 0%
\end{array}%
\right] ,|i|=2.\text{ }  \notag \\
&&
\end{eqnarray}%
The gauge transformation $G=\mathbbm{1}\oplus U_{1}\oplus \left(
U_{1}U_{2}\right) $ can be used to rewrite them in the standard forms. After
substituting $a^{[i]}$ for $a_{0}^{[i]}$ if $|i|=0$, $a_{0}^{i}U_{1}^{%
\dagger }$ if $|i|=1$, and $a_{0}^{i}U_{2}^{\dagger }U_{1}^{\dagger }$ if $%
|i|=2$, we can express the local matrices into more compact form: 
\begin{equation}
A^{[i]}=Y_{3}^{|i|}\otimes a^{[i]}.
\end{equation}

To obtain the irreducible MPS, it should be guaranteed that $A^{[i]}$ must
have no irreducible invariant subspace commuting with $Q_{3}$. If there was
such an invariant subspace corresponding to the projector $\tilde{P}$, it
should have the form $\tilde{P}=\text{diag}(\tilde{P}_{0},\tilde{P}_{1},%
\tilde{P}_{2})$. According to $A^{[i]}\tilde{P}=\tilde{P}A^{[i]}\tilde{P}$,
it further satisfies 
\begin{equation}
a^{[i]}\tilde{P}_{0}=\tilde{P}_{0}a^{[i]}\tilde{P}_{0},a^{[i]}\tilde{P}_{1}=%
\tilde{P}_{1}a^{[i]}\tilde{P}_{1},a^{[i]}\tilde{P}_{2}=\tilde{P}_{2}a^{[i]}%
\tilde{P}_{2},
\end{equation}%
for $\forall |i|=0$. To exclude such a situation, we must impose the
necessary condition that the \textquotedblleft charge-0\textquotedblright\
subalgebra spanned by all $\{a^{[i_{1}]}\cdots a^{[i_{p}]}\}$ with $\forall
p\in \mathbb{N}$ and $\sum_{l=1}^{p}|i_{l}|=0$ is simple. In the following,
when we mention the \textquotedblleft charge-0\textquotedblright\
subalgebra, it has the same definition, but it is the simple matrix algebra
with different dimension.

So the local matrices $A^{[i]}$ are irreducible if $A^{[i]}$ can be gauge
transformed into $Y_{3}^{|i|}\otimes a^{[i]}$ and the \textquotedblleft
charge-0\textquotedblright\ sub-algebra is a simple matrix algebra. These
conditions imply that $A^{[i]}$ span a non-trivial simple $\mathbb{Z}_{3}$
graded algebra. The graded center consists of multiples of $\mathbbm{1}%
,Y_{3} $, and $Y_{3}^{2}$. Taking the trivial simple $\mathbb{Z}_{3}$ graded
algebra of MPS into consideration, we obtain the conclusion that a $\mathbb{Z%
}_{3}$ parafermion MPS is irreducible iff $A^{[i]}$ span a simple $\mathbb{Z}%
_{3}$ graded algebra.

\subsection{Topological order in $\mathbb{Z}_{3}$ parafermion MPS}

The characteristic properties of the parafermionic MPS can be found in the
transfer matrix 
\begin{equation}
\mathbb{E}=\sum_{i}A^{[i]}\otimes \bar{A}^{[i]}.
\end{equation}%
For the trivial algebra MPS, the irreducible matrices $A^{[i]}$ span a
simple algebra. The corresponding transfer matrix forms a completely
positive map\cite{CPmap}, whose eigenvalue spectrum is real and
non-negative, and the largest eigenvalue is non-degenerate.

However, for the non-trivial type algebra MPS, the transfer matrix can be
expressed as 
\begin{equation}
\mathbb{E}=\sum_{i}\left[ Y_{3}^{|i|}\otimes a^{[i]}\right] \otimes \left[
Y_{3}^{|i|}\otimes \bar{a}^{[i]}\right] .
\end{equation}%
By supposing $\tilde{\sigma}_{R}$ as the right eigenvector of the sub-block
transfer matrix $\tilde{\mathbb{E}}=\sum_{i}a^{[i]}\otimes \bar{a}^{[i]}$
with the real eigenvalue $\lambda $, i.e. $\sum_{i}a^{[i]}\tilde{\sigma}%
_{R}a^{[i]\dagger }=\lambda \tilde{\sigma}_{R}$, it can be easily verified
that $\sigma _{R,j}=Y_{3}^{|j|}\otimes \tilde{\sigma}_{R}$ with $|j|=0,1,2$
are three eigenvectors of the transfer matrix $\mathbb{E}$ with the same
eigenvalue $\lambda $. It can be further proved that all eigenvalues of the
transfer matrix $\mathbb{E}$ have at least three-fold degeneracy. The
details are given in the Sec. IV.B. The largest eigenvalue and the
corresponding eigenvectors stem from the sub-block transfer matrix $\tilde{%
\mathbb{E}}$, so the three-fold degeneracy of the transfer matrix spectrum
reflects the existence of unpaired parafermion edge zero modes,
characterizing the topological order in one dimension. In contrast, the
largest eigenvalue of the transfer matrix of a symmetry protected
topological state is non-degenerate.

\begin{figure}[tbp]
\centering
\includegraphics[width=8.7cm]{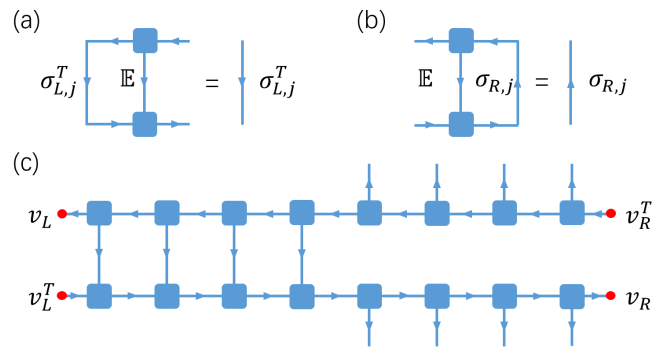}\newline
\caption{(a) The left dominant eigenvectors of the transfer matrix $\mathbb{E%
}$ give rise to the left fixed points $\protect\sigma _{L,j}=Y_{3}^{|j|}%
\otimes \tilde{\protect\sigma}_{L}$. (b) The right dominant eigenvectors
produce the right fixed points $\protect\sigma _{R,j}=Y_{3}^{|j|}\otimes 
\tilde{\protect\sigma}_{R}$ (c) Bipartition of the MPS via partially
contracting the physical degrees of freedom. The red dots are the vectors
fixing the boundaries of MPS.}
\label{figure1}
\end{figure}

Moreover, according to the holographic principle, the left and right
dominant eigenvectors of the transfer operator determine the reduced density
matrix in the thermodynamic limit\cite{Bultinck2017,ESofPEPS}. We can study
the entanglement spectrum via a bipartition of the parafermionic MPS. Here
we merely consider the non-trivial MPS. Supposing that the left and right
dominant eigenvectors of the sub-block transfer matrix $\tilde{\mathbb{E}}$
are given by $\tilde{\sigma}_{L}$ and $\tilde{\sigma}_{R}$, the transfer
matrix $\mathbb{E}$ has three left and three right dominant eigenvectors $%
\sigma _{L,j}=Y_{3}^{|j|}\otimes \tilde{\sigma}_{L}$ and $\sigma
_{R,j}=Y_{3}^{|j|}\otimes \tilde{\sigma}_{R}$, respectively, displayed in
Fig. \ref{figure1}(a) and Fig. \ref{figure1}(b). Notice that the tensors $%
v_{R}\otimes v_{R}^{T}$ and $v_{L}^{T}\otimes v_{L}$ fixing the double layer
tensor network must be charge zero, where $v_{R}$ and $v_{L}$ are the right
and left boundary vectors of MPS, respectively, as shown in \ref{figure1}%
(c). The entanglement Hamiltonian $H_{E}$ in the thermodynamic limit is thus
determined by the left and right charge zero fixed-points as 
\begin{equation}
e^{H_{E}}=(\mathbbm{1}\otimes \tilde{\sigma}_{L}^{\ast })(\mathbbm{1}\otimes 
\tilde{\sigma}_{R})=\mathbbm{1}\otimes \tilde{\sigma}_{L}^{\ast }\tilde{%
\sigma}_{R},
\end{equation}%
Hence the entanglement spectrum have at least three-fold degeneracy, fully
determined by the structure of $A^{[i]}$. 

To summarize, there are only two types of irreducible $\mathbb{Z}_{3}$
parafermionic MPS. One type corresponds to the local matrices spanning a
trivial simple $\mathbb{Z}_{3}$ graded algebra, so the dominant eigenvector
of the transfer matrix is unique and the entanglement spectrum is not
necessarily degenerate. The other type corresponds to the non-trivial simple 
$\mathbb{Z}_{3}$ graded algebra spanned by local matrices. The full transfer
matrix spectrum has at least three-fold degenerate eigenvalues, and so does
the entanglement spectrum in the thermodynamic limit. The degeneracy of the
transfer matrix spectrum implies the existence of the unpaired $\mathbb{Z}%
_{3}$ parafermion zero modes. Actually such analysis can be generalized to
all $\mathbb{Z}_{p}$ parafermionic MPS for topological phases, and the
necessary degeneracy of the transfer matrix spectrum as well as the
degeneracy of the entanglement spectrum in thermodynamic limit just depend
on the structure of $A^{[i]}$.

\subsection{Irreducibility of $\mathbb{Z}_{4}$ parafermion MPS}

Beside the topological and the trivial phases, there is a spontaneous
symmetry breaking phase in the classification of $\mathbb{Z}_{4}$
parafermion chains. The non-trivial orthogonal projectors of the invariant
subspaces are still given by $P_{i}=Q_{4}^{i}P_{0}Q_{4}^{-i}$ with $%
i=0,1,2,3 $, where each of $P_{i}$ is associated with an irreducible
invariant subspace. The discussion is divided into three different
situations.

1. $\left[ P_{0},Q_{4}\right] =0$

Since the irreducible invariant subspace projector $P_{0}$ contains four
different charge sectors, the forms of the local matrices $A^{[i]}$ are
given by Eq. (\ref{block}), and the irreducible $A^{[i]}$ span a trivial
simple $\mathbb{Z}_{4}$ graded algebra. Both the transfer matrix spectrum
and the entanglement spectrum are not necessarily degenerate.

2. $\left[ P_{0},Q_{4}\right] \neq 0$ but $\left[ P_{0},Q_{4}^{2}\right] =0$

In this case we have only two orthogonal projectors $P_{0}$ and $P_{1}$ for
two irreducible invariant subspaces. According to $%
Q_{4}P_{0}Q_{4}^{-1}+P_{0}=$ $\mathbbm{1}$ and $P_{0}^{2}=P_{0}$, $P_{0}$
can be determined as 
\begin{equation}
P_{0}=\frac{1}{2}\left[ 
\begin{array}{cccc}
\mathbbm{1} & 0 & U_{1} & 0 \\ 
0 & \mathbbm{1} & 0 & U_{2} \\ 
U_{1}^{\dagger } & 0 & \mathbbm{1} & 0 \\ 
0 & U_{2}^{\dagger } & 0 & \mathbbm{1}%
\end{array}%
\right] ,
\end{equation}%
where $U_{1}$ and $U_{2}$ are unitary block matrices and both the charge-0
and charge-1 sectors have the same dimension as the charge-2 and charge-3
sectors. The relation $A^{[i]}P_{j}=P_{j}A^{[i]}P_{j}$ and the gauge
transformation $G=\mathbbm{1}\oplus \mathbbm{1}\oplus U_{1}\oplus U_{2}$
lead to 
\begin{equation}
A^{[i]}=\text{diag}\left(
a_{0}^{[i]},a_{1}^{[i]},a_{0}^{[i]},a_{1}^{[i]}\right) \times \left(
Y_{4}^{|i|}\otimes \mathbbm{1}\right)
\end{equation}%
where we have substituted $a_{1}^{[i]}$ for $a_{1}^{[i]}U_{1}^{\dagger }$ if 
$|i|=1$, $a_{0}^{[i]}$ for $a_{0}^{[i]}U_{1}^{\dagger }$, $a_{1}^{[i]}$ for $%
a_{1}^{[i]}U_{2}^{\dagger }$ if $|i|=2$, and $a_{0}^{[i]}$ for $%
a_{0}^{[i]}U_{2}^{\dagger }$ if $|i|=3$. It is further required that the
dimensions of the four charge sectors are the same. Then, by permuting the
order of basis, $(0,1,2,3)\rightarrow (0,2,1,3)$, these local matrices can
display an even $\mathbb{Z}_{2}$ graded structure: 
\begin{eqnarray}
A^{[i]} &=&\left[ 
\begin{array}{cccc}
a_{0}^{[i]} & 0 & 0 & 0 \\ 
0 & a_{0}^{[i]} & 0 & 0 \\ 
0 & 0 & a_{1}^{[i]} & 0 \\ 
0 & 0 & 0 & a_{1}^{[i]}%
\end{array}%
\right] ,|i|=0;  \notag  \label{Z4matrices} \\
A^{[i]} &=&\left[ 
\begin{array}{cccc}
0 & 0 & a_{0}^{[i]} & 0 \\ 
0 & 0 & 0 & a_{0}^{[i]} \\ 
0 & a_{1}^{[i]} & 0 & 0 \\ 
a_{1}^{[i]} & 0 & 0 & 0%
\end{array}%
\right] ,|i|=1;  \notag \\
A^{[i]} &=&\left[ 
\begin{array}{cccc}
0 & a_{0}^{[i]} & 0 & 0 \\ 
a_{0}^{[i]} & 0 & 0 & 0 \\ 
0 & 0 & 0 & a_{1}^{[i]} \\ 
0 & 0 & a_{1}^{[i]} & 0%
\end{array}%
\right] ,|i|=2;  \notag \\
A^{i} &=&\left[ 
\begin{array}{cccc}
0 & 0 & 0 & a_{0}^{[i]} \\ 
0 & 0 & a_{0}^{[i]} & 0 \\ 
a_{1}^{[i]} & 0 & 0 & 0 \\ 
0 & a_{1}^{[i]} & 0 & 0%
\end{array}%
\right] ,|i|=3.
\end{eqnarray}

Since we have $\left[ P_{0},Q_{4}^{2}\right] =0$, the invariant subspace $%
P_{0}$ contains the even and odd $\mathbb{Z}_{2}$ parity sectors. In
addition, $A^{[i]}$ must not contain an invariant subspace whose projector
commutes with the operator $Q_{4}$. Similar to the $\mathbb{Z}_{3}$ case, it
is necessary that the \textquotedblleft charge-zero\textquotedblright\
subalgebras of matrix algebras spanned by $a_{0}^{[i]}$ and $a_{1}^{[i]}$
are simple, respectively. Then the local matrices $A^{[i]}$ span a simple
algebra in the $\mathbb{Z}_{4}$ graded sense, but it is semisimple in the $%
\mathbb{Z}_{2}$ graded sense\cite{Fidkowski2011}. Because the semisimple
algebra can split into irreducible ones, we can reduce the local matrices
and break the $\mathbb{Z}_{4}$ symmetry down to $\mathbb{Z}_{2}$ symmetry.
To make it explicit, another gauge transformation 
\begin{equation}
G^{\prime }=\frac{1}{\sqrt{2}}\left[ 
\begin{array}{cccc}
0 & \mathbbm{1} & 0 & -\mathbbm{1} \\ 
i\mathbbm{1} & 0 & -i\mathbbm{1} & 0 \\ 
0 & \mathbbm{1} & 0 & \mathbbm{1} \\ 
\mathbbm{1} & 0 & \mathbbm{1} & 0%
\end{array}%
\right] ,
\end{equation}%
changes the local matrices into the canonical form: 
\begin{eqnarray}
A^{[i]} &=&\omega _{4}^{|i|}d^{[i]}\oplus d^{[i]},  \notag \\
d^{[i]} &=&\left[ 
\begin{array}{cc}
a_{1}^{[i]} & 0 \\ 
0 & a_{0}^{[i]}%
\end{array}%
\right] ,|i|=0,2;  \notag \\
d^{[i]} &=&\left[ 
\begin{array}{cc}
0 & a_{1}^{[i]} \\ 
a_{0}^{[i]} & 0%
\end{array}%
\right] ,|i|=1,3.
\end{eqnarray}%
The block diagonal form of $A^{[i]}$ represents a spontaneous symmetry
breaking phase\cite{Chen-Gu-Wen-2011}. One may question why it is not a
topological phase? The answer is that it is impossible to transform $A^{[i]}$
into $Y_{2}^{|i|}\otimes d^{[i]}$, which is required in the topological
phase. The reducing process actually mixes the charge-0 and charge-2
sectors, as well as the charge-1 and charge-3 sectors. In this sense, the $%
\mathbb{Z}_{4}$ symmetry breaks down to the $\mathbb{Z}_{2}$ symmetry. After
reducing, the sub-block matrices $d^{[i]}$ span an even type simple $\mathbb{%
Z}_{2}$ graded algebra, shown in Eq. (\ref{evenZ2}). So there is no Majorana
zero edge modes. The two degenerate ground states are purely resulted from
the spontaneous symmetry breaking, and can be transformed into each other
via the representation of the $\mathbb{Z}_{4}$ charge symmetry generator $%
U_{4}=\text{diag}(\mathbbm{1},\omega \mathbbm{1},\omega ^{2}\mathbbm{1}%
,\omega ^{3}\mathbbm{1})$. Therefore, the two parts in the direct sum are
connected by $\sum_{i}U_{ij}d^{[j]}=\omega ^{|i|}d^{[i]}$. Moreover, the
parafermions $\chi _{i}$ can form bosons $\chi _{i}^{2}$ and its
anti-particles $\chi _{i}^{3}$. But the Majorana fermions can not be
obtained from $\mathbb{Z}_{4}$ parafermions, so there is no non-trivial
topological phase with Majorana edge zero modes in this classification.

3. $\left[ P_{0},Q_{4}\right] \neq 0$ and $\left[ P,Q_{4}^{2}\right] \neq 0$

Following the general procedure of deriving the irreducible MPS, the
standard form of the local matrices can be written as 
\begin{equation}
A^{[i]}=Y_{4}^{|i|}\otimes a^{[i]},|i|=0,1,2,3,
\end{equation}%
where the \textquotedblleft charge-zero\textquotedblright\ sub-algebra is a
simple matrix algebra and the local matrices span a non-trivial simple $%
\mathbb{Z}_{4}$ graded algebra with the non-trivial center consisting of
multiples of $\mathbbm{1}$, $Y_{4}$, $Y_{4}^{2}$ and $Y_{4}^{3}$. So the
minimal four-fold degeneracy of the transfer matrix spectrum and the
entanglement spectrum can be found, implying the existence of $\mathbb{Z}%
_{4} $ parafermion edge zero modes. Thus, this kind of irreducible MPS
corresponds to the non-trivial topological phase.

\subsection{Classification of $\mathbb{Z}_{6}$ parafermion MPS}

In this case, there exist more than one topological phases, exhibiting a
richer physics than $\mathbb{Z}_{3}$ and $\mathbb{Z}_{4}$ parafermions. From
the irreducibility of $\mathbb{Z}_{6}$ parafermion MPS, we first assume that
there is an irreducible invariant subspace projector $P_{0}$, and then
divide the discussion into four different situations.

1. $\left[ P_{0},Q_{6}\right] =0$

This means that the irreducible invariant subspace $P_{0}$ contains all six
charge sectors. The matrices in the invariant subspace $P_{0}$ span the
trivial simple $\mathbb{Z}_{6}$ graded algebra, corresponding to the trivial
phase.

2. $\left[ P_{0},Q_{6}\right] \neq 0$ but $\left[ P_{0},Q_{6}^{2}\right] =0$

The above two relations automatically lead to $\left[ P_{0},Q_{6}^{3}\right]
\neq 0$. Considering the constraints $P_{0}+Q_{6}P_{0}Q_{6}^{-1}=\mathbbm{1}$
and $P_{0}^{2}=P_{0}$, we can express 
\begin{equation}
P_{0}=\frac{1}{2}\left[ 
\begin{array}{cccccc}
\mathbbm{1} & 0 & 0 & U_{1} & 0 & 0 \\ 
0 & \mathbbm{1} & 0 & 0 & U_{2} & 0 \\ 
0 & 0 & \mathbbm{1} & 0 & 0 & U_{3} \\ 
U_{1}^{\dagger } & 0 & 0 & \mathbbm{1} & 0 & 0 \\ 
0 & U_{2}^{\dagger } & 0 & 0 & \mathbbm{1} & 0 \\ 
0 & 0 & U_{3}^{\dagger } & 0 & 0 & \mathbbm{1}%
\end{array}%
\right] ,
\end{equation}%
where $U_{1}$, $U_{2}$ and $U_{3}$ are unitary matrices required by the
idemponency. Applying $P_{0}$ to $A^{[i]}$, we obtain 
\begin{equation}
A^{[i]}=\text{diag}\left(
a_{0}^{[i]},a_{1}^{[i]},a_{2}^{[i]},a_{0}^{[i]},a_{1}^{[i]},a_{2}^{[i]}%
\right) \times \left( Y_{6}^{|i|}\otimes \mathbbm{1}\right)
\end{equation}%
with some redefinitions. Via permuting the basis $(0,1,2,3,4,5,6)\rightarrow
(0,2,4,1,3,5)$, we can rewrite $A^{[i]}$ into the standard form on the $%
\mathbb{Z}_{2}$ parity basis: 
\begin{eqnarray}
A^{[i]} &=&\left[ 
\begin{array}{cc}
d^{[i]} & 0 \\ 
0 & U^{\dagger }d^{[i]}U%
\end{array}%
\right] ,|i|=0,2,4;  \notag \\
A^{[i]} &=&\left[ 
\begin{array}{cc}
0 & d^{[i]} \\ 
U^{\dagger }d^{[i]}U^{\dagger } & 0%
\end{array}%
\right] ,|i|=1,3,5,
\end{eqnarray}%
where 
\begin{eqnarray}
U &=&\left[ 
\begin{array}{ccc}
0 & \mathbbm{1} & 0 \\ 
0 & 0 & \mathbbm{1} \\ 
\mathbbm{1} & 0 & 0%
\end{array}%
\right] ,  \notag \\
d^{[i]} &=&\left[ 
\begin{array}{ccc}
a_{0}^{[i]} & 0 & 0 \\ 
0 & a_{2}^{[i]} & 0 \\ 
0 & 0 & a_{1}^{[i]}%
\end{array}%
\right] ,|i|=0,1;  \notag \\
d^{[i]} &=&\left[ 
\begin{array}{ccc}
0 & a_{0}^{[i]} & 0 \\ 
0 & 0 & a_{2}^{[i]} \\ 
a_{1}^{[i]} & 0 & 0%
\end{array}%
\right] ,|i|=2,3;  \notag \\
d^{[i]} &=&\left[ 
\begin{array}{ccc}
0 & 0 & a_{0}^{[i]} \\ 
a_{2}^{[i]} & 0 & 0 \\ 
0 & a_{1}^{[i]} & 0%
\end{array}%
\right] ,|i|=4,5.
\end{eqnarray}%
Note that the dimensions of the six charge sectors must be the same. Such a
form of $A^{[i]}$ actually satisfies the odd type $\mathbb{Z}_{2}$ graded
algebra shown in Eq. (\ref{oddZ2}), so these matrices can be further
transformed via the gauge transformation $\mathbb{I}\oplus U$ into the
standard form 
\begin{equation}
A^{[i]}=Y_{2}^{|i|}\otimes d^{[i]}.  \label{6-2}
\end{equation}%
Since $[P_{0},Q_{6}^{2}]=0$ means that $P_{0}$ containing three $\mathbb{Z}%
_{3}$ charge sectors is irreducible, $d^{[i]}$ span a trivial simple $%
\mathbb{Z}_{3}$ graded algebra. Moreover, it should be emphasized that there
exists no invariant subspace whose projector commutes with the operator $%
Q_{6}$. To guarantee such a situation, all the \textquotedblleft
charge-zero\textquotedblright\ subalgebras of matrix algebras spanned by $%
a_{0}^{[i]}$, $a_{1}^{[i]}$ and $a_{2}^{[i]}$ should be simple,
respectively. This determines $A^{[i]}$ to span a type of the simple $%
\mathbb{Z}_{6}$ graded algebra, which is the similar to the odd-type $%
\mathbb{Z}_{2}$ graded algebra. The graded center of this non-trivial
algebra consists of multiples of $\mathbbm{1}$ and $Y_{2}$.

Because the $\mathbb{Z}_{6}$ symmetry can not be broken for $\mathbb{Z}_{6}$
parafermions, the $\mathbb{Z}_{6}$ graded structure can not be removed, and
the local matrices are not reduced. From Eq. (\ref{6-2}), we can show that
the full transfer matrix spectrum have two-fold degeneracy, implying that
the entanglement spectrum is at least two-fold degeneracy in the
thermodynamic limit and there exist unpaired Majorana zero edge modes. Thus,
this phase is a $\mathbb{Z}_{6}$ symmetric non-trivial topological phase
without any symmetry breaking, but it shares the same property as the $%
\mathbb{Z}_{2}$ non-trivial topological phase.

3. $\left[ P_{0},Q_{6}\right] \neq 0$ and $\left[ P_{0},Q_{6}^{2}\right]
\neq 0$ but $\left[ P_{0},Q_{6}^{3}\right] =0$

Similar to the above case, the standard form of $A^{[i]}$ can be written as 
\begin{eqnarray}
A^{[i]} &=&Y_{3}^{|i|}\otimes d^{[i]},  \notag  \label{6-3} \\
d^{[i]} &=&\left[ 
\begin{array}{cc}
a_{0}^{[i]} & 0 \\ 
0 & a_{1}^{[i]}%
\end{array}%
\right] ,|i|=0,1,2;  \notag \\
d^{[i]} &=&\left[ 
\begin{array}{cc}
0 & a_{0}^{[i]} \\ 
a_{1}^{[i]} & 0%
\end{array}%
\right] ,|i|=3,4,5.
\end{eqnarray}%
Provided that the \textquotedblleft charge-zero\textquotedblright\
sub-algebras spanned by $a_{0}^{[i]}$ and $a_{1}^{[i]}$ are simple, $A^{[i]}$
will span a type of simple $\mathbb{Z}_{6}$ graded algebra, which is the
same as the non-trivial simple $\mathbb{Z}_{3}$ graded algebra with the
non-trivial center. Eq.(\ref{6-3}) determines that the eigenvalue spectrum
of the transfer matrix have at least three-fold degeneracy, indicating that
there are three-fold degenerate entanglement spectrum in the thermodynamic
limit and unpaired $\mathbb{Z}_{3}$ parafermion edge zero modes. Thus, the
resulting MPS represents another $\mathbb{Z}_{6}$ symmetric non-trivial
topological phase with the same property as the $\mathbb{Z}_{3}$ non-trivial
topological phase.

4. $\left[ P_{0},Q_{6}\right] \neq 0$, $\left[ P_{0},Q_{6}^{2}\right] \neq 0$
and $\left[ P_{0},Q_{6}^{3}\right] \neq 0$

In this case, the form of matrices $A^{[i]}$ can be expressed as the
standard form 
\begin{equation}
A^{[i]}=Y_{6}^{|i|}\otimes a^{[i]},
\end{equation}%
which span another type of non-trivial simple $\mathbb{Z}_{6}$ graded
algebra provided the \textquotedblleft charge zero\textquotedblright\
sub-algebra spanned by $a^{[i]}$ is simple. The graded center consists of
multiples of the regular representation of $\mathbb{Z}_{6}$ group. Because
of the six-fold degeneracy of the transfer matrix spectrum, this case
corresponds to a topological non-trivial phase with $\mathbb{Z}_{6}$
parafermion edge zero modes, yielding six-fold degeneracy of the
entanglement spectrum in the thermodynamic limit.

To summarize, there exist three topologically distinct phases and one
trivial phase. Their local matrices of the irreducible MPS form different
types of simple $\mathbb{Z}_{6}$ graded algebra with the distinct centers.
These topological phases are characterized by the Majorana zero modes, $%
\mathbb{Z}_{3}$ parafermion zero modes, and $\mathbb{Z}_{6}$ parafermion
zeros modes at the edges of systems, respectively.

\subsection{Classification of $\mathbb{Z}_{8}$ parafermion MPS}

The reason why we are interested in this case is that it is alleged that
there exist two distinct spontaneous symmetry breaking phases from the $%
\mathbb{Z}_{8}$ to $\mathbb{Z}_{4}$ symmetry: one has a two-fold ground
state degeneracy purely due to the symmetry breaking, and another has a
four-fold ground state degeneracy resulting from both spontaneous symmetry
breaking and topological order\cite{Quella}. Here we can carefully examine
these results from the irreducibility perspective of the MPS. To gain more
information about related phases, especially the symmetry breaking phases,
we divide our discussion into four situations.

1. $\left[ P_{0},Q_{8}\right] =0$

The matrices associated to the irreducible invariant subspace projector $%
P_{0}$ contain eight $\mathbb{Z}_{8}$ charge sectors, corresponding a
trivial simple $\mathbb{Z}_{8}$ graded algebra and thus a trivial phase.

2. $\left[ P_{0},Q_{8}\right] \neq 0$ but $\left[ P_{0},Q_{8}^{2}\right] =0$

It implies that the invariant subspace denoted by the projector $P_{0}$ does
not contain all eight $\mathbb{Z}_{8}$ charge sectors, but it contains only
four $\mathbb{Z}_{4}$ charge sectors. The commutation relation $%
[P_{0},Q_{8}^{4}]=0$ indicates that $\mathbb{Z}_{2}$ parity sectors are also
contained in $P_{0}$ as well. Then the resulting MPS is trivial from both $%
\mathbb{Z}_{4}$ and $\mathbb{Z}_{2}$ symmetry point of view. Since the $%
P_{0} $ invariant space is irreducible, we can not break the symmetry down
to $\mathbb{Z}_{2}$. The local matrices span a type of simple $\mathbb{Z}_{8}
$ graded algebra which is the same as the trivial semisimple $\mathbb{Z}_{4}$
graded algebra, provided that the charge-0 subalgebras are simple. Hence it
is reducible and the $\mathbb{Z}_{8}$ symmetry spontaneously breaks down to $%
\mathbb{Z}_{4}$, attributing a two-fold degeneracy. Because the trivial
graded algebra has no nontrivial graded center, the corresponding MPS
describes a pure symmetry breaking phase. Actually the spontaneous symmetry
breaking is related to the phenomenon of boson condensation, since we have $%
[\chi _{i}^{4},\chi _{j}^{4}]=0$.

3. $\left[ P_{0},Q_{8}\right] \neq 0$ and $\left[ P_{0},Q_{8}^{2}\right]
\neq 0$ but $\left[ P_{0},Q_{8}^{4}\right] =0$

Then the invariant subspace denoted by the projector $P_{0}$ contains
neither eight $\mathbb{Z}_{8}$ charge sectors nor four $\mathbb{Z}_{4}$
charge sectors, but it only contains two $\mathbb{Z}_{2}$ charge sectors,
leading to a trivial MPS from the $\mathbb{Z}_{2}$ symmetry point of view.
Since $P_{0}^{2}=P_{0}$ and 
\begin{equation}
P_{0}+Q_{8}P_{0}Q_{8}^{-1}+Q_{8}^{2}P_{0}Q_{8}^{-2}+Q_{8}^{3}P_{0}Q_{8}^{-3}=%
\mathbbm{1},
\end{equation}%
we can find that $P_{0}$ is 
\begin{equation}
\frac{1}{4}\left[ 
\begin{array}{cccccccc}
\mathbbm{1} & 0 & U_{1} & 0 & U_{2} & 0 & U_{3} & 0 \\ 
0 & \mathbbm{1} & 0 & U_{4} & 0 & U_{5} & 0 & U_{6} \\ 
U_{1}^{\dagger } & 0 & \mathbbm{1} & 0 & U_{1}^{\dagger }U_{2} & 0 & 
U_{1}^{\dagger }U_{3} & 0 \\ 
0 & U_{4}^{\dagger } & 0 & \mathbbm{1} & 0 & U_{4}^{\dagger }U_{5} & 0 & 
U_{4}^{\dagger }U_{6} \\ 
U_{2}^{\dagger } & 0 & U_{1}^{\dagger }U_{2} & 0 & \mathbbm{1} & 0 & 
U_{2}^{\dagger }U_{3} & 0 \\ 
0 & U_{5}^{\dagger } & 0 & U_{5}^{\dagger }U_{4} & 0 & \mathbbm{1} & 0 & 
U_{5}^{\dagger }U_{6} \\ 
U_{3}^{\dagger } & 0 & U_{3}^{\dagger }U_{1} & 0 & U_{3}^{\dagger }U_{2} & 0
& \mathbbm{1} & 0 \\ 
0 & U_{6}^{\dagger } & 0 & U_{6}^{\dagger }U_{4} & 0 & U_{6}^{\dagger }U_{5}
& 0 & \mathbbm{1}%
\end{array}%
\right] ,  \notag
\end{equation}%
where $U_{i}$ are unitary matrices with the same dimension. According to $%
A^{[i]}P_{j}=P_{j}A^{[i]}P_{j}$ and with some proper substitutions, we can
express the local matrices as 
\begin{equation}
A^{[i]}=\text{diag}%
(a_{0}^{[i]},a_{1}^{[i]},a_{0}^{[i]},a_{1}^{[i]},a_{0}^{[i]},a_{1}^{[i]},a_{0}^{[i]},a_{1}^{[i]})\times (Y_{8}^{|i|}\otimes %
\mathbbm{1}).
\end{equation}%
By permuting the basis, $(0,1,2,3,4,5,6,7)\rightarrow (0,2,4,6,1,3,5,7)$, we
can explicitly show that the matrices $A^{[i]}$ have an even $\mathbb{Z}_{2}$
graded structure as%
\begin{eqnarray}
A^{[i]} &=&\left[ 
\begin{array}{cc}
Y_{4}^{|i|/2}\otimes a_{0}^{[i]} & 0 \\ 
0 & Y_{4}^{|i|/2}\otimes a_{1}^{[i]}%
\end{array}%
\right] ,|i|=0,2,4,6;  \notag  \label{40} \\
A^{[i]} &=&\left[ 
\begin{array}{cc}
0 & Y_{4}^{(|i|-1)/2}\otimes a_{0}^{[i]} \\ 
Y_{4}^{(|i|+1)/2}\otimes a_{1}^{[i]} & 0%
\end{array}%
\right] ,|i|=1,3,5,7,  \notag \\
&&
\end{eqnarray}%
These matrices span a simple $\mathbb{Z}_{8}$ graded algebra, which is the
same as a trivial semisimple $\mathbb{Z}_{2}$ graded algebra, provided that
the \textquotedblleft charge-zero\textquotedblright\ sub-algebras are
simple. From above equation one can exclude the possibility of existence of
topological order. Since there are no non-trivial graded center but four
irreducible invariant spaces, we can reduce the simple $\mathbb{Z}_{8}$
graded algebra to the even-type simple $\mathbb{Z}_{2}$ graded algebras, and
the $\mathbb{Z}_{8}$ symmetry is broken down to the $\mathbb{Z}_{2}$
symmetry, attributing four-fold degenerate ground states. Then Eq. (\ref{40}%
) can be transformed into a direct sum of four sets of even simple algebra $%
\mathbb{Z}_{2}$ matrices via the gauge transformation 
\begin{equation}
G=\frac{1}{2}\left[ 
\begin{array}{cccccccc}
0 & 0 & 0 & 0 & -\mathbbm{1} & \mathbbm{1} & -\mathbbm{1} & \mathbbm{1} \\ 
-i\mathbbm{1} & i\mathbbm{1} & -i\mathbbm{1} & i\mathbbm{1} & 0 & 0 & 0 & 0
\\ 
0 & 0 & 0 & 0 & \mathbbm{1} & -i\mathbbm{1} & -\mathbbm{1} & i\mathbbm{1} \\ 
\omega _{8}\mathbbm{1} & \omega _{8}^{7}\mathbbm{1} & \omega _{8}^{5}%
\mathbbm{1} & \omega _{8}^{3}\mathbbm{1} & 0 & 0 & 0 & 0 \\ 
0 & 0 & 0 & 0 & \mathbbm{1} & i\mathbbm{1} & -\mathbbm{1} & -i\mathbbm{1} \\ 
\omega _{8}^{3}\mathbbm{1} & \omega _{8}^{5}\mathbbm{1} & \omega _{8}^{7}%
\mathbbm{1} & \omega _{8}\mathbbm{1} & 0 & 0 & 0 & 0 \\ 
0 & 0 & 0 & 0 & -\mathbbm{1} & -\mathbbm{1} & -\mathbbm{1} & -\mathbbm{1} \\ 
-\mathbbm{1} & -\mathbbm{1} & -\mathbbm{1} & -\mathbbm{1} & 0 & 0 & 0 & 0%
\end{array}%
\right] ,
\end{equation}%
which mixes the even $\mathbb{Z}_{8}$ charge sectors $(0,2,4,6)$ as well as
the odd $\mathbb{Z}_{8}$ charge sectors $(1,3,5,7)$, separately. After this
transformation, the local matrices are rewritten as 
\begin{eqnarray}
A^{[i]} &=&\left( \omega _{8}^{2|i|}d^{[i]}\right) \oplus \left( \omega
_{8}^{|i|}d^{[i]}\right) \oplus \left( {\omega _{8}}^{3|i|}d^{[i]}\right)
\oplus d^{[i]},  \notag \\
d^{[i]} &=&\left[ 
\begin{array}{cc}
a_{1}^{[i]} & 0 \\ 
0 & a_{0}^{[i]}%
\end{array}%
\right] ,|i|=0,2,4,6;  \notag \\
d^{[i]} &=&\left[ 
\begin{array}{cc}
0 & a_{1}^{[i]} \\ 
a_{0}^{[i]} & 0%
\end{array}%
\right] ,|i|=1,3,5,7.
\end{eqnarray}%
Notice that the four-fold degeneracy is only contributed by the spontaneous
symmetry breaking, and there exists no topological order.

4. $\left[ P_{0},Q_{8}\right] \neq 0$, $\left[ P_{0},Q_{8}^{2}\right] \neq 0$
and $\left[ P_{0},Q_{8}^{4}\right] \neq 0$

In this case, the irreducible invariant subspace $P_{0}$ only contains one
charge sector. The irreducible matrices is given by 
\begin{equation}
A^{[i]}=Y_{8}^{|i|}\otimes a^{[i]},
\end{equation}%
which span a non-trivial simple $\mathbb{Z}_{8}$ graded algebra with the
non-trivial center consisting of multiples of regular representation of $%
\mathbb{Z}_{8}$, provided the \textquotedblleft
charge-zero\textquotedblright\ sub-algebra is a simple algebra. The transfer
matrix spectrum has eight-fold degeneracy and entanglement spectrum in the
thermodynamic limit should have at least eight-fold degeneracy,
corresponding to the $\mathbb{Z}_{8}$ symmetric topological phase. Hence
there is only one topological phase.

\section{General Results for $\mathbb{Z}_{p}$ parafermion MPS}

\subsection{Irreducibility and classification}

Summarizing the above several examples, we can obtain the general
classification for all $\mathbb{Z}_{p}$ parafermion phases. First we assume
an irreducible invariant subspace projector $P_{0}$ for all matrices $%
A^{[i]} $. Then we consider the commutation relations between $Q_{p}^{r}$
and $P_{0}$, where $r$ is the divisor of $p$. It can be proved that the
number of different cases denoted by commutation relations between $%
Q_{p}^{r} $ and $P_{0}$ is the number of divisor $p$. Each case is labelled
by the smallest divisor $n\in \{r\}$ such that $[Q_{p}^{n},P_{0}]=0$.
Together with the idempotency constrain, the structures of all irreducible
invariant subspace projectors as well as that of $A^{[i]}$ can be determined.

Actually, Eq. (\ref{standradform}) can be written in a more concise form: 
\begin{equation}
A^{[i]}=\text{diag}\left( a_{0}^{[i]},a_{1}^{[i]},\cdots
,a_{p-1}^{[i]}\right) \times \left( Y_{p}^{|i|}\otimes \mathbbm{1}\right) .
\label{trivial}
\end{equation}%
For the case $[P_{0},Q_{p}]=0$, $a_{s}^{[i]}$ with $s\in \mathbb{Z}_{p}$ are
not equal for all $i$ under gauge transformations and redefinitions, and all 
$A^{[i]}$ span the trivial simple $\mathbb{Z}_{p}$ graded algebra. The MPS
generated by the matrices of Eq. (\ref{trivial}) belong to the trivial phase.

However, for the case $[P_{0},Q_{p}^{r}]\neq 0$ with $r<n$ and $%
[P_{0},Q_{p}^{n}]=0$, the relation $a_{s}^{[i]}=a_{(s+p/n)\bmod p}^{[i]}$
satisfies under gauge transformations and redefinitions. There exist $p/n$
unequal sub-block matrices $a_{s}^{[i]}$ ($s\in \mathbb{Z}_{p/n}$). Then,
there are two different situations, depending on whether $n$ and $p/n$ are
mutually prime or not. If $n$ and $p/n$ are mutually prime, by using a
charge-preserving gauge transformation represented by a permutation matrix,
the local matrices can be transformed into 
\begin{eqnarray}
A^{[i]} &=&Y_{n}^{|i|}\otimes d^{[i]},  \notag  \label{non-trivial2} \\
d^{[i]} &=&\text{diag}\left( a_{0}^{[i]},\cdots ,a_{p/n-1}^{[i]}\right)
\times \left( Y_{p/n}^{|i|}\otimes \mathbbm{1}\right) .
\end{eqnarray}%
Under the condition that the \textquotedblleft charge-0\textquotedblright\
sub-algebras are simple, all $A^{[i]}$ are irreducible and span a
non-trivial simple $\mathbb{Z}_{p}$ graded algebra with a non-trivial
center. The MPS generated by Eq.(\ref{non-trivial2}) indicate a $\mathbb{Z}%
_{p}$ symmetric topological phase with unpaired $\mathbb{Z}_{n}$ parafermion
zero edge modes. This topological phase is characterized by the $n$-fold
degenerate transfer matrix spectrum and entanglement spectrum.

In the case where $n$ and $p/n$ are not mutually prime, it is impossible
that the local matrices can be transformed into the form of Eq. (\ref%
{non-trivial2}). The reason is that we can write $Y_{p}\sim Y_{p/n}\otimes
Y_{n}$ only if $n$ and $p/n$ are mutually prime. Actually, we can express $%
Y_{p}\sim \tilde{Q}_{p/n}\otimes Y_{n}$, where $\tilde{Q}_{p/n}=\text{diag}%
\left( 1,\omega _{p}^{1},\omega _{p}^{2},\cdots ,\omega _{p}^{n-1}\right) $,
so the local matrices can be transformed via a gauge transformation into 
\begin{equation}
A^{[i]}=\tilde{Q}_{p/n}^{|i|}\otimes d^{[i]}.  \label{SSBform}
\end{equation}%
However, the gauge transformation breaks the $\mathbb{Z}_{p}$ charge
symmetry but preserves the $\mathbb{Z}_{p/n}$ charge symmetry. Provided that
the \textquotedblleft charge-0\textquotedblright\ sub-algebras are simple,
all $A^{[i]}$ span a simple $\mathbb{Z}_{p}$ graded algebra, which is the
same as the trivial semisimple $\mathbb{Z}_{p/n}$ graded algebra. And it can
be reduced into the trivial simple $\mathbb{Z}_{p/n}$ graded algebra. So
this situation corresponds to the phases where the $\mathbb{Z}_{p}$ symmetry
is spontaneously broken down to $\mathbb{Z}_{p/n}$.

So a conclusion can be drawn that the number of phases is equal to the
number of divisors of $p$, and every divisor $n$ uniquely labels a different
gapped phase\cite{Pollmann2013,Quella}. The topological phases including the
trivial phase are labeled by $n$ satisfying that $n$ and $p/n$ are mutually
prime. The different parafermion gapped phases have one-to-one
correspondence to the different $\mathbb{Z}_{p}$ graded algebras. A more
concise summary is shown in Table. \ref{table2}. 
\begin{table*}[tbp]
\caption{Comparison of different types of gapped phases of $\mathbb{Z}_{p}$
parafermions characterized by the integer $n$.}
\label{table2}%
\begin{tabular}{|C{2.5cm}|C{4.5cm}|C{5cm}|C{4.5cm}|}
  \hline
  \textbf{Phase} & \textbf{Topological} & \textbf{Symmetry breaking} & \textbf{Trivial}\\ \hline
  Label $n$ & $n$ and $p/n$ are coprime & $n$ and $p/n$ are not coprime & $n=1$ \\ \hline
Transfer matrix spectrum&  $n$-fold degenerate spectrum&  $n$-fold degenerate real part of spectrum(for whole ground space)& Non-degenerate largest eigenvalue \\ \hline
Algebra& Non-trivial simple $\mathbb{Z}_{p}$ graded& Trivial semisimple $\mathbb{Z}_{p/n}$ graded& Trivial simple $\mathbb{Z}_{p}$ graded \\ \hline
\end{tabular}
\end{table*}

\subsection{Degeneracy of transfer matrix spectrum}

For the symmetry breaking phases, the degenerate ground states can be
transformed with each other by acting the $\mathbb{Z}_{p}$ symmetry
generator $U=\text{diag}\left( \mathbb{I},\omega \mathbbm{1},\omega ^{2}%
\mathbbm{1},\text{...},\omega ^{p-1}\mathbbm{1}\right) $ several times,
i.e., $\sum_{i}U_{ij}d^{[j]}=\omega ^{|i|}d^{[i]}$. If we act the $\mathbb{Z}%
_{p}$ charge operator $n$ times, we will go back to the original state,
since it is $\mathbb{Z}_{p/n}$ symmetric. Thus the local matrices is shown
in Eq.(\ref{SSBform}). Therefore, the transfer matrices for the whole ground
state subspace of symmetry breaking phases are given by 
\begin{equation}
\mathbb{E}=\sum_{i}\left[ \left( \bigoplus\limits_{r=0}^{n-1}\omega
_{p}^{|i|r}d^{[i]}\right) \otimes \left( \bigoplus\limits_{r=0}^{n-1}\bar{%
\omega}_{p}^{|i|r}\bar{d}^{[i]}\right) \right] .
\end{equation}

Actually, because all $Y_{n}^{r}$ with $r\in \mathbb{Z}_{n}$ can be
diagonalized simultaneously, $A^{[i]}=Y_{n}^{|i|}\otimes d^{[i]}\sim
Q_{n}^{|i|}\otimes d^{[i]}$, the transfer matrices for topological phases
can be transformed into 
\begin{equation}
\mathbb{E}^{\prime }=\sum_{i}\left[ \left(
\bigoplus\limits_{r=0}^{n-1}\omega _{n}^{|i|r}d^{[i]}\right) \otimes \left(
\bigoplus\limits_{r=0}^{n-1}\bar{\omega}_{n}^{|i|r}\bar{d}^{[i]}\right) %
\right] .
\end{equation}%
These two expressions are very similar but the phase factors are different.
Their eigenvalue spectra are equivalent to those of the following matrices%
\cite{Chen-Gu-Wen-2011} 
\begin{equation}
\bigoplus\limits_{r,r^{\prime }=0}^{n-1}\sum_{i}\omega _{p}^{|i|(r-r^{\prime
})}d^{[i]}\otimes \bar{d}^{[i]},\text{ }\bigoplus\limits_{r,r^{\prime
}=0}^{n-1}\sum_{i}\omega _{n}^{|i|(r-r^{\prime })}d^{[i]}\otimes \bar{d}%
^{[i]}.
\end{equation}%
When $r=r^{\prime }$, $\widetilde{\mathbb{E}}\left( r,r\right)
=\sum_{i}d^{[i]}\otimes \bar{d}^{[i]}$ defines a block transfer matrix,
whose largest eigenvalue is non-degenerate and the spectrum is real and
non-negative. Thus the real eigenvalues are $n$-fold degenerate for both
cases. On the other hand, $\widetilde{\mathbb{E}}(r,r^{\prime
})=\sum_{i}\omega _{p}^{|i|(r-r^{\prime })}d^{[i]}\otimes \bar{d}^{[i]}$ and 
$\widetilde{\mathbb{E}}^{\prime }\left( r,r^{\prime }\right) =\sum_{i}\omega
_{n}^{|i|(r-r^{\prime })}d^{[i]}\otimes \bar{d}^{[i]}$ for $r\neq r^{\prime }
$ denote mixed transfer matrices\cite{SSB-MPS}, and their eigenvalues are
complex, and the magnitudes of eigenvalues are smaller than unity\cite%
{MPSrep}.

However, for topological phases, taking the advantage of the fact that $r\in 
\mathbb{Z}_{n}$ and the periodicity of $\omega _{n}$ is also $n$, there are $%
n$ possible values of $\{r,r^{\prime }\}$ for a fixed value of $r-r^{\prime }
$. While for the symmetry breaking phases, $\widetilde{\mathbb{E}}^{\prime
}(r,r^{\prime })$ does not have such a property. Therefore, the complex
eigenvalues of the transfer matrix are also $n$-fold degenerate for
topological phases, while the complex eigenvalues of the transfer matrix of
the symmetry breaking phase are not necessarily $n$-fold degenerate.
Different from those features displayed in the spontaneous symmetry breaking
phases, the degeneracy of the full transfer matrix spectrum is the unique
characteristic property for topological phases. This degeneracy is a clear
evidence of the existence of parafermion zero edge modes. In contrast, the
degeneracy of entanglement spectrum can appear for both the topological
order phases and the symmetry protected topological phases.

\subsection{Understanding topological order in MPS formalism}

It is known that all one-dimensional bosonic gapped systems can support
short-range entanglement without any intrinsic topological order. However,
the one-dimensional fermion and parafermion systems can probably have the
topological order. Actually, since the statistics is not well-defined in one
dimension, the parafermion chains only emerge at the edges of a
two-dimensional fractionalized topological states\cite{Lindner,Clarke,Cheng}%
, and the topological order is inherited from the bulk of fractional
topological insulators. So these phases are distinct from the symmetry
protected topological phases, and it is more appropriate to recognize them
as invertible topological order\cite{zoo}.

Previously such a topological order is characterized by strong zero edge
modes. A strong zero edge mode carrying the $\mathbb{Z}_{p}$ charge is
defined by an operator localized at the edges, which commutes with the model
Hamiltonian\cite{Fendley2012,Majorana doubling}. But such a strong zero edge
mode can be easily washed away when an arbitrary small perturbation is
introduced into the fixed point model Hamiltonian\cite%
{Fendley2012,Jermyn,Fernando}. Nevertheless, even in the absence of the
strong zero edge modes, the gapped phases still display topological nature,
and the weak edge modes commuting with the ground state subspace exist. It
is more proper describing the topological order from the ground state wave
functions rather than the model Hamiltonians.

In the view point of the fermionic/parafermionic MPS, we have understood
that different topological phases correspond to the non-trivial simple $%
\mathbb{Z}_{p}$ graded algebras with different non-trivial centers. The
matrix $Y_{n}$ features the non-trivial graded structure and acts as the
fractionalized charge operators, characterizing the topological order. In
fact the matrix $Y_{n}$ has more profound indications. It can also be
regarded as the gauge symmetry of the local tensors, namely, $(Y_{n}\otimes %
\mathbbm{1})A^{[i]}(Y_{n}\otimes \mathbbm{1})^{-1}=A^{[i]}$, which plays a
crucial role and becomes the necessary condition of the topological order.
It can be further verified that the non-trivial algebra MPS are the
G-injective MPS\cite{PEPS-Deg-Topo}. Actually, the properties of the
topological order in fermionic/parafermionic MPS are similar to those found
in PEPS in two dimensional systems. We believe that our formalism provides
the proper way to describe the topological order in one dimension.

\section{Conclusion and Outlook}

Using the graded tensor product, we have encoded the parafermion statistics
into the Fock space, and identified it as a graded vector space. Then, based
on the Fock space, we have constructed the general MPS for all
one-dimensional gapped phases for $\mathbb{Z}_{p}$ parafermions without
extra symmetry. We have also investigated several specific examples,
covering all possible gapped phases. From the analysis of irreducibility of
MPS, it has been found that all parafermion gapped phases can be classified
by MPS. By identifying algebras spanned by the irreducible local matrices,
we also find different phases have one-to one correspondence to different
simple $\mathbb{Z}_{p}$ graded algebras. We have further analyzed the
properties of the corresponding transfer matrix spectra and entanglement
spectra. The topological phases can be identified by the unique property
that the full transfer matrix spectra are $n$-fold degenerate ($n\leq p$).

Our formalism can be easily generalized to the classification with extra
symmetries, including the general on-site symmetries\cite{Pollmann2013} and
time-reversal symmetry\cite{Meidan}. Moreover, some exact solvable models
can also be designed within our MPS formalism and the characteristic
properties of various phases can be more easily calculated\cite{Fernando}.
And the renormalization group of fermionic/parafermionic MPS can be
developed as well. Finally, our present formulation may be useful to
investigate the fermionic/parafermionic PEPS in more than one spatial
dimension\cite{Gu-Verstraete-Wen,Eisert,VerstraeteMPO}. Thus the present
theoretical framework and forthcoming results will greatly enrich our
understanding of low-dimensional strongly correlated many-body systems.

\textit{Acknowledgment.- }The authors would like to thank Guo-Yi Zhu and
Zi-Qi Wang for their stimulating discussion and acknowledges the support of
National Key Research and Development Program of China (No.2017YFA0302902).

\begin{center}
\textbf{Appendix: Some useful concepts}\\[0pt]
\end{center}

In the content of our paper, we frequently mention the graded algebra,
irreducibility, simplicity and semisimplicity. In this appendix, we will
list the definitions of these concepts, which are summarized from the
literature\cite{Clliford,MPDO,Fidkowski2011}.

\textbf{Graded algebra}. An algebra $\mathcal{A}$ is said to be $\mathbb{Z}%
_{p}$ graded if there is a decomposition of the underlying vector space $%
\mathcal{A}=\oplus _{n=0}^{p}\mathcal{A}^{n}$ such that $\mathcal{A}^{n}%
\mathcal{A}^{m}=\mathcal{A}^{n+m}$. If $A\in \mathcal{A}^{n}$, then $A$ is
said to be homogeneous of degree $n$.

\textbf{Irreducibility}. If the local tensor has a block upper triangular
form:. 
\begin{equation}
A^{[i]}=\left( 
\begin{array}{cc}
a_{1}^{[i]} & a_{0}^{[i]} \\ 
0 & a_{2}^{[i]}%
\end{array}%
\right) .
\end{equation}%
The MPS generated by $A^{[i]}$ doesn't depend on $a_{0}^{[i]}$. In fact,
there exist a subspace $S_{1}$ which is invariant under the action, $%
A^{[i]}S_{1}\in S_{1}$. Then we can choose $a_{0}^{[i]}=0$ and assume $S_{1}$
doesn't contain any other invariant subspace. Denoting $P_{1}(Q_{1}=%
\mathbbm{1}-P_{1})$ as the orthogonal projector onto $S_{1}(S_{1}^{\perp })$%
, we have 
\begin{equation}
A^{[i]}P_{1}=P_{1}A^{[i]}P_{1},\quad Q_{1}A^{[i]}=Q_{1}A^{[i]}Q_{1}.
\end{equation}%
Then $P_{1}A^{[i]}P_{1}$ generates the irreducible MPS. Actually, any MPS
can be gauge-transformed into a direct sum of irreducible MPS. For
parafermion systems, the irreducibility is different. The invariant subspace
must be a graded vector space, projected by $P_{1}$ commutating with charge
matrix. If it contain other subspaces which is not a graded vector space, we
can not reduce it.

\textbf{Simplicity and semisimplicity}. The simplicity and semisimplicity
are usually defined in terms of (finite-dimensional) representations, i.e.,
the vector spaces over the field on which the algebra acts linearly. An
algebra is called semisimple if any representation of a subgroup has a
complementary representation. Any representation of a semisimple algebra
splits into irreducible ones. A semisimple algebra is called simple if it
has a unique irreducible representation. Any semisimple algebra is a direct
sum of simple algebras. For example, the group algebra splits as a direct
sum of algebra spanned by its irreducible representations. Actually, the
algebra spanned by local matrices for topological and symmetry breaking
phases is the simple $\mathbb{Z}_{p}$ graded algebra, but they are
semisimple without $\mathbb{Z}_{p}$ grading, corresponding to the
interpretation for phases of clock spin chains.

\end{document}